\documentclass[prl,showpacs,showkeys,nofootinbib,twocolumn]{revtex4}

\usepackage{graphicx}  %
\usepackage{amsmath}
\usepackage{bm}  %

\newcommand{\beq}{\begin{equation}}
\newcommand{\eeq}{\end{equation}}
\newcommand{\bqa}{\begin{eqnarray}}
\newcommand{\eqa}{\end{eqnarray}}

\def\mqo2{{\!\!\!}}

\begin{document}

\title{Resonant Dimer Relaxation in Cold Atoms with a Large Scattering Length}

\author{Eric Braaten}
\affiliation{Department of Physics,
         The Ohio State University, Columbus, OH\ 43210, USA}

\author{H.-W. Hammer}
\affiliation{Helmholtz-Institut f{\"u}r Strahlen- und Kernphysik 
   (Theorie), Universit{\"a}t Bonn, 53115 Bonn, Germany}

\date{January 29, 2007}

\begin{abstract}
Efimov physics refers to universal phenomena associated with a 
discrete scaling symmetry in the 3-body problem with a large 
scattering length.  The first experimental evidence for 
Efimov physics was the recent observation of a resonant peak 
in the 3-body recombination rate for $^{133}$Cs atoms
with large negative scattering length. 
There can also be resonant peaks in the atom-dimer relaxation rate 
for large positive scattering length.
We calculate the atom-dimer relaxation rate
as a function of temperature and show how measurements of the 
relaxation rate can be used to determine accurately the parameters 
that govern Efimov physics.   
\end{abstract}

\smallskip
\maketitle

The {\it Efimov effect} is a remarkable phenomenon that can occur in a 3-body
system of nonrelativistic particles with short-range interactions  
when the S-wave scattering length $a$ is tuned to $\pm \infty$. 
Vitaly Efimov discovered in 1970 that in this limit there can be 
infinitely many 3-body bound states with an accumulation point at the 
scattering threshold and a geometric spectrum \cite{Efimov70}: 
each successive bound state is shallower than the previous one 
by a multiplicative factor.  In the case of identical bosons 
with mass $m$, 
this factor is $e^{2\pi/s_0} \approx 515$, where $s_0 \approx 1.00624$. 
The binding energies of the {\it Efimov states} for $a = \pm \infty$ 
can be expressed as 
\begin{eqnarray}
E^{(n)}_T = (e^{2\pi/s_0})^{n_*-n} \hbar^2 \kappa^2_* /m,
\label{kappa-star}
\end{eqnarray}
where $\kappa_*$ is the binding wavenumber of the branch of Efimov states 
labeled by $n_*$.  The geometric spectrum is a signature of a 
{\it discrete scaling symmetry}
with {\it discrete scaling factor} $e^{\pi/s_0}\approx 22.7$.

The Efimov effect is just one example of the universal
phenomena characterized by a discrete scaling symmetry that govern the
3-body system in the {\it scaling limit} in which the range of the 
interaction is negligible compared to $|a|$ \cite{Efimov71,Efimov79}.
For a review of these phenomena, 
which we refer to as {\it Efimov physics}, see Ref.~\cite{Braaten:2004rn}.
For simplicity, we focus on the case of identical bosons.  
We refer to the particles as {\it atoms}, 
their 2-body bound states as {\it dimers}, 
and their 3-body bound states as {\it trimers}.  
If the scattering length $a$ is large and positive, there is a  
{\it shallow dimer} with binding energy $E_D=\hbar^2/(ma^2)$.
Efimov pointed out that there is an infinite sequence of
negative values of $a$ for which there is an Efimov trimer 
at the 3-atom scattering threshold \cite{Efimov71}: 
$a=(e^{\pi/s_0})^n a_*'$, where $a_*' \approx  - 1.56 \kappa_*^{-1}$
\cite{BH02}. 
There is also an infinite sequence of
positive values of $a$ for which there is an Efimov trimer at the 
atom-dimer scattering threshold \cite{Efimov71}:
$a=(e^{\pi/s_0})^n a_*$, where $a_* \approx 0.0708 \kappa_*^{-1}$
\cite{BH02}. 
Another example of Efimov physics is an infinite sequence
of positive values of $a$ for which the 3-body recombination rate 
into the shallow dimer vanishes \cite{NM-99,EGB-99,BBH-00}. 
The universal aspects of Efimov physics are 
determined by two parameters: the scattering length $a$
and the Efimov parameter $\kappa_*$.

The alkali atoms used in most cold atom experiments 
have many deeply-bound diatomic molecules ({\it deep dimers}).
Efimov physics is modified by the existence of the deep dimers,
because Efimov trimers can decay into an atom and a deep dimer.
Deep dimers also provide inelastic scattering channels
for three atoms or for an atom and a dimer. If there is an Efimov trimer 
near the scattering threshold, it can give  
a resonant enhancement of inelastic scattering processes.
When $a \approx (e^{\pi/s_0})^n a_*'$, the resonant inelastic process 
is 3-atom recombination into a deep dimer \cite{EGB-99,BH01}.
When $a \approx (e^{\pi/s_0})^n a_*$, the resonant inelastic process 
is {\it dimer relaxation}, in which the collision of an atom 
and a shallow dimer produces an atom and a deep
dimer \cite{Braaten:2003yc}.
If there are deep dimers, the universal phenomena associated 
with Efimov physics are determined by three parameters:
$a$, $\kappa_*$, and a parameter $\eta_*$ that determines 
the widths of Efimov trimers \cite{Braaten:2003yc}.

Finally, more than 35 years after Efimov's discovery, 
the first experimental evidence for Efimov physics 
has emerged.  In a beautiful experiment with cold $^{133}$Cs atoms, 
the Innsbruck group has observed a resonance in the 3-body 
recombination rate that can be attributed to an Efimov trimer 
near the 3-atom threshold \cite{Grimm06}. 
They used the magnetic field to control the scattering length of 
$^{133}$Cs atoms in the lowest hyperfine state by exploiting a 
Feshbach resonance at $-12$ G. 
Since inelastic 2-body losses are energetically forbidden, 
the dominant loss mechanism is 3-body recombination.
By varying the magnetic field from 0 to 150 G, they were able to change the
scattering length from $-2500 \; a_0$ through 0 to $+1600 \; a_0$,
where $a_0$ is the Bohr radius.
They observed a giant loss feature at $a \approx -850 \; a_0$. 
At the lowest temperature at which they measured the loss rate 
as a function of $a$, which was 10nK, it can be fit 
rather well by the universal formula of 
Refs.~\cite{Braaten:2004rn,Braaten:2003yc} with $\eta_* = 0.06(1)$.

The Innsbruck group also measured the 3-body recombination rate for
positive values of $a$ reached by increasing the magnetic field 
through the zero of the scattering length \cite{Grimm06}. 
They observed a local minimum in the inelastic loss rate at 
$a \approx 210 \; a_0$.
The minimum could be due to Efimov physics, although the value of 
$a$ is not large compared to the van der Waals length scale
for Cs atoms: $(mC_6/\hbar^2)^{1/4} \approx 200 \; a_0$.  Another 
complication was that the lowest 
temperature that was reached for positive $a$ was 200 nK.
Thermal effects can be taken into account if the 3-body recombination 
rate is known as a function of the collision energy.
The universal predictions for nonzero energy
have not yet been calculated. 

One way to confirm the discovery of Efimov physics 
would be to observe a second loss feature related to the first feature 
by the discrete scaling symmetry.  In the case of identical bosons, 
the large size of the discrete scaling factor $e^{\pi/s_0} \approx 22.7$
makes this difficult. Since the 3-body recombination rate scales 
like $a^4$, the loss rate at the second loss feature is larger than
that at the first feature by a factor of $22.7^4 \approx 2 \times 10^5$.
Another way is to observe the correlations 
between different aspects of Efimov physics that are predicted by 
universality.  This requires the accurate determination of the 
parameters $\kappa_*$ and $\eta_*$ that govern Efimov physics.
It may also require the calculation of the universal predictions 
at nonzero temperature to account properly for thermal effects.
The effects of temperature on the resonant inelastic loss 
for negative $a$ have been considered in 
Refs.~\cite{dInc04,Jonsell06,YFT06}.

For positive values of $a$, the Innsbruck group has also created mixtures
of $^{133}$Cs atoms and shallow dimers \cite{FB18Santos}, 
which has allowed them to study atom-dimer collisions.  
The most dramatic manifestation of Efimov physics for $a>0$
is the resonant enhancement of dimer relaxation for $a$ near
 $(e^{\pi/s_0})^n a_*$. 
The Innsbruck group has observed a resonance in inelastic atom-dimer
collisions near $a \approx 400 \; a_0$ \cite{FB18Santos}. 
In this paper, we calculate the rate for inelastic atom-dimer collisions 
in a cold gas of atoms and dimers as a function of the scattering length
and the temperature.  By measuring the position and 
line shape of the resonance
and using our results to take into account the temperature, 
it should be possible to determine the Efimov parameters 
$\kappa_*$ and $\eta_*$ accurately.  Once these parameters are determined, 
they can be used to make accurate predictions for other universal 
phenomena associated with Efimov physics.
 
The collision energy for an atom and a shallow dimer is
$E = 3 \hbar^2 k^2/(4m)$, where $k$ is the wavenumber of the atom 
or dimer in the center-of-mass frame.
The differential cross section 
for elastic atom-dimer scattering  is
\begin{eqnarray}
\frac{d\sigma_{AD}}{d\Omega}=\left| f_{AD}(k,\theta)\right|^2 \,,
\label{dsigma-AD}
\end{eqnarray}
where $f_{AD}(k,\theta)$ is the scattering amplitude, which has the
partial wave expansion
\begin{eqnarray}
f_{AD}(k,\theta) = \sum_{L=0}^\infty {2L+1 \over k \cot \delta^{AD}_L (k) -ik}
 P_L(\cos \theta) \,.
\label{fk-AD}
\end{eqnarray}
The elastic cross section $\sigma_{AD}(E)$ is obtained by integrating 
over the solid angle of $4\pi$.
The leading terms in the expansion of $k \cot \delta^{AD}_0(k)$ 
in powers of $k$ define the {\it atom-dimer scattering length} $a_{AD}$ 
and the {\it atom-dimer effective range} $r_{s,AD}$:
\begin{eqnarray}
k \cot \delta^{AD}_0(k) = - 1/a_{AD} + \mbox{$1 \over 2$} r_{s,AD} k^2 +
\ldots \,.
\label{rAD-def}
\end{eqnarray}
The optical theorem relates the total cross section,
which is the sum of the elastic and inelastic cross sections, 
to the $\theta \to 0$ limit of the 
elastic scattering amplitude in Eq.~(\ref{fk-AD}):
\begin{eqnarray}
\sigma_{AD}^{\rm (total)}(E)  = \frac{4\pi}{k} \, {\rm Im} \, 
f_{AD}(k,\theta = 0) \,.
\label{optical-AD}
\end{eqnarray}
If there are no deep dimers,
the scattering is elastic up to the dimer-breakup
threshold at $E=E_D$.

We first consider the case in which there are no deep dimers,
so the phase shifts $\delta^{AD}_L(k)$ 
are real-valued below the dimer-breakup threshold.
The phase shifts for $L \ge 1$ are universal functions of $ka$ only.  
The $L = 0$ phase shift is sensitive to 3-body interactions at
short distances and therefore depends also on $a \kappa_*$.
Efimov used unitarity to derive powerful constraints on the 
dependence on $a \kappa_*$ that he referred to as the {\it radial law}. 
He used the radial law to deduce an analytic expression for
the atom-dimer scattering length $a_{AD}$ \cite{Efimov79}
up to numerical constants that were first calculated by 
Simenog and Sinitchenko \cite{Sim81}. They
have been calculated more accurately
using an effective field theory \cite{BHK99,BHK99b,BH02}. 
This method was used in Ref.~\cite{BH02} to calculate the atom-dimer 
effective range $r_{s,AD}$. These universal results are
\begin{eqnarray}
a_{AD} &=& \big( 1.46 + 2.15 \cot [s_0 \ln (a/a_*)] \big) \; a \,,
\label{aAD-univ}
\\
r_{s,AD} &=&
\big( 1.13 + 0.73 \cot[s_0 \ln(a/a_*) +0.98 ] \big)^2 a \,.
\nonumber
\end{eqnarray}
Numerical calculations indicate that $r_{s,AD}$
vanishes within the numerical accuracy
when $a \approx (e^{\pi/s_0})^n 4.85 \ a_*$ \cite{BH02}.
The above expression for $r_{s,AD}$ is compatible with Efimov's radial law
and also satisfies this additional constraint.

We now consider the case in which there are deep dimers,
so dimer relaxation provides
inelastic atom-dimer scattering channels.
The large  binding energy of the deep dimer is released through the  
large kinetic energies of the recoiling atom and dimer. 
The event rate $\beta$ for {\it dimer relaxation} 
is defined so that the
number of dimer relaxation events per time and per volume 
in a gas of atoms with number density $n_A$ 
and dimers with number density $n_D$ is $\beta n_A n_D$.  
If the atom and the deep dimer from the relaxation 
process have such large kinetic energies that they escape 
from the system, the rate of decrease in the number densities is 
\begin{eqnarray}
{d \ \over d t} n_A & = &  {d \ \over d t} n_D
= - \beta n_A n_D \,.
\label{dn-deact}
\end{eqnarray}
The event rate $\beta$ can be expressed in terms of a 
statistical average of the inelastic atom-dimer cross section:
\begin{eqnarray}
\beta = {3 \hbar \over 2 m}
\left\langle k \ \sigma_{AD}^{\rm (inelastic)}(E) \right\rangle \,.
\label{beta-T}
\end{eqnarray}
The inelastic cross section is the difference between the total 
cross section in Eq.~(\ref{optical-AD}) and the elastic cross section
obtained by integrating Eq.~(\ref{dsigma-AD}) over angles.
If the temperature is large compared to the critical temperature for
Bose-Einstein condensation, the thermal average
indicated by the angular brackets in Eq.~(\ref{beta-T})
can be carried out using Boltzmann statistics.
In a gas consisting of atoms and dimers in thermal equilibrium
(but not chemical equilibrium), the Boltzmann factor is 
$\exp(-3 \hbar^2 k^2/(4 m k_B T))$.

In Ref.~\cite{Braaten:2003yc}, 
we pointed out that in the scaling (or zero-range) limit, 
the cumulative effect of all the deep dimers on Efimov physics 
can be taken into account rigorously through one
additional parameter $\eta_*$. 
The effects of Efimov physics can be dramatic only if 
$\eta_*$ is much less than 1. If the universal 
expression for a scattering amplitude for the case of no deep dimers 
is known as an analytic function of $\kappa_*$ or $a_*$, 
the corresponding result for a system with deep
dimers can be obtained without any additional calculation simply
by making the substitution   
\begin{eqnarray}
\ln a_* \longrightarrow \ln a_* - i \eta_*/s_0 \,.
\label{logsub}
\end{eqnarray}
Since the $L=0$ atom-dimer phase shift is the only one that depends on
$\kappa_*$, the $L=0$ channel is the only one that contributes to
dimer relaxation at leading order.
Making the substitution in Eq.~(\ref{logsub})
in Eqs.~(\ref{aAD-univ}),
we obtain complex-valued expressions for the atom-dimer scattering length 
and effective range:
\begin{eqnarray}
a_{AD} &=& \big( 1.46 + 2.15 \cot [s_0 \ln (a/a_*) + i \eta_*] \big) \; a \,,
\label{aAD-complex}
\\
r_{s,AD} &=&
\big( 1.13 + 0.73 \cot[s_0 \ln(a/a_*) +0.98 + i \eta_*] \big)^2 a \,.
\nonumber
\end{eqnarray}
If the expansion for $k \cot \delta^{AD}_0(k)$ in 
Eq.~(\ref{rAD-def}) is truncated after the effective range term, 
the inelastic cross section reduces to
\begin{eqnarray}
\sigma_{AD}^{\rm (inelastic)}(E)  &=& 
\frac{4\pi(-{\rm Im} \, a_{AD} -{1\over2} |a_{AD}|^2 {\rm Im}\, r_{s,AD} k^2)}
{k \left| 1 + i a_{AD} k - {1\over2} a_{AD} r_{s,AD} k^2 \right|^2}\,.
\nonumber
\\
\label{ksig-inelastic}
\end{eqnarray}
Our truncation of the effective range expansion in Eq.~(\ref{rAD-def})
limits the validity of Eq.~(\ref{ksig-inelastic}) 
to collision energies $E \ll E_D$,
where $E_D = \hbar^2/(ma^2)$.  This expression could be extended
straightforwardly to the entire region $E < E_D$ by using a 
parameterization of the atom-dimer phase shift below the dimer-breakup 
threshold given in Ref.~\cite{BH02}. 
The rate constant $\beta$ for $T \ll E_D$ can be obtained by 
inserting the cross section in Eq.~(\ref{ksig-inelastic})
into Eq.~(\ref{beta-T}).
In the low-temperature limit, $\beta$
reduces to $6 \pi \hbar(-{\rm Im} \, a_{AD})/m$,
which can be expressed as \cite{Braaten:2003yc}
\begin{eqnarray}
\beta &=& 
{20.3 \sinh(2\eta_*) \over \sin^2 [s_0 \ln (a/a_*)] + \sinh^2 \eta_*} 
\; {\hbar a \over m}  \,.
\label{beta:a>0}
\end{eqnarray}
If $\eta_*$ is small, the maximum value of $\beta$ occurs when
$a \approx [1 + \sinh^2 \eta_*/(2 s_0^2)] a_*$,
which is $a \approx 1.0018 \ a_*$ if $\eta_* = 0.06$.

\begin{figure}[htb]
\centerline{\includegraphics*[width=8.5cm,angle=0,clip=true]{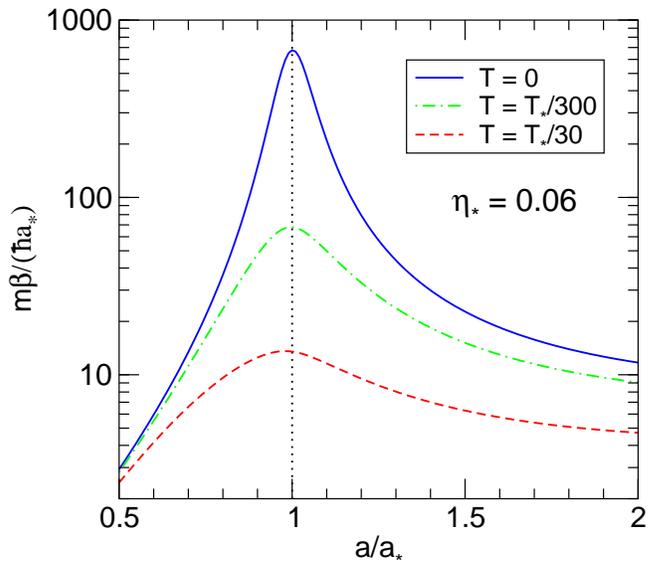}}
\vspace*{0.0cm}
\caption{The dimer relaxation length $m\beta/\hbar$ (in units of $a_*$)
as a function of $a/a_*$ for $\eta_*=0.06$. The
curves are for several  values of the temperature: $T = 0$ (solid line),
$T_*/300$ (dot-dashed line), $T_*/30$ (dashed line),
where $k_B T_* = \hbar^2/(m a_*^2)$. }
\label{fig:beta-T}
\end{figure}

\begin{figure}[htb]
\centerline{\includegraphics*[width=8.5cm,angle=0,clip=true]
{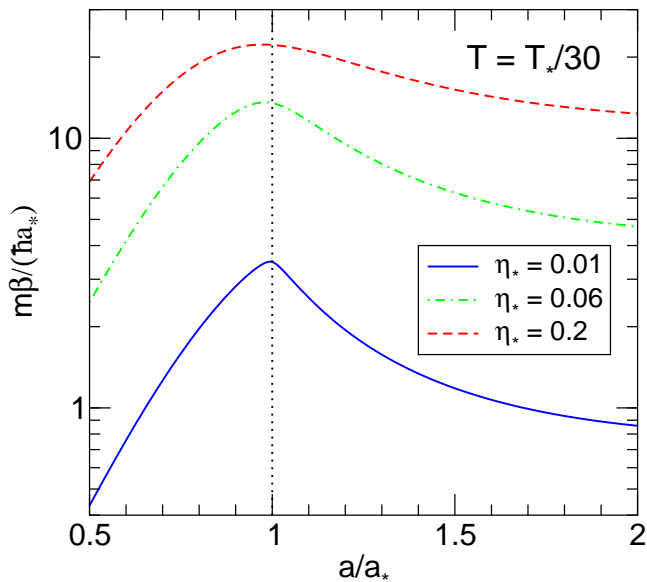}}
\vspace*{0.0cm}
\caption{The dimer relaxation length $m\beta/\hbar$ (in units of $a_*$)
as a function of $a/a_*$ for $T=T_*/30$, where $k_B T_* = \hbar^2/(m a_*^2)$.
The curves are for $\eta_* = 0.01$ (solid line),
0.06 (dot-dashed line), and 0.2 (dashed line). }
\label{fig:beta-eta}
\end{figure}

The resonant line shape for the dimer relaxation rate $\beta$ 
as a function of $a$ is illustrated in Figs.~\ref{fig:beta-T} and
\ref{fig:beta-eta}.
In Fig.~\ref{fig:beta-T}, we show the dimer relaxation length 
$m \beta/\hbar$ as a function of $a/a_*$
for $\eta_* = 0.06$ and several values of $T$. 
The temperature is given in units of $T_* = \hbar^2/(k_B m a_*^2)$,
which is the dimer binding energy for $a = a_*$.
These results can be applied to any atomic system with large 
positive $a$
by using the appropriate value of $a_*$ to convert to physical units.
In Fig.~\ref{fig:beta-eta}, we show $m \beta/\hbar$ as a function 
of $a/a_*$ for $T = T_*/30$ and several values of $\eta_*$.
As $T$ increases, 
the height of the resonant peak decreases and its width increases,
but the location of the peak remains very close to $a_*$.
If $\eta_*$ is increased,
the height of the resonance peak and its width both increase.
The discrete scaling symmetry implies that $\beta$ has similar peaks
at $a \approx (22.7)^n a_*$, $n=1,2,\ldots$
with maximum values of $\beta$ that are larger by $(22.7)^n$.
The fact that the peak remains near $a_*$
implies that $a_*$ (and therefore the Efimov parameter $\kappa_*$)
can be determined accurately by measurements at nonzero temperature 
without any need for thermal corrections.  
The value of $\eta_*$ can be 
determined from the line shape of $\beta$ as a function of $a$,
if the temperature is known to sufficient accuracy.

The resonance in inelastic atom-dimer scattering observed 
by the Innsbruck group occurs when the scattering length is 
$a \approx 400 \; a_0$ \cite{FB18Santos}.  
Since the  van der Waals length scale for Cs atoms
is $(mC_6/\hbar^2)^{1/4} \approx 200 \; a_0$, this value of $a$ 
is not deep into the universal region.  If we ignore this problem, 
we can apply our universal results to $^{133}$Cs atoms in the 
vicinity of the resonance by setting $a_* = 400 \; a_0$.  
Using $\hbar^2/(k_B m) = 1.3032$ K$a_0^2$ for Cs atoms, 
we find that the temperature scale set by $a_*$ is $T_* = 8145$ nK.
The temperatures in Fig.~\ref{fig:beta-T} 
correspond to 0, 27.2 nK, and 272 nK. The temperature
in Fig.~\ref{fig:beta-eta} corresponds to 272 nK.  
The measurements of $\beta$ by the Innsbruck group were carried out 
at temperatures near 250 nK \cite{FB18Santos}.  
A reasonable guess for $\eta_*$ is the value 0.06 that was measured
for large negative $a$ on the other side of the zero of $a$.
The value for $\eta_*$ for this region of large positive $a$ 
could be measured by fitting 
the line shape to our universal results, provided the temperature 
can be determined with sufficient accuracy. 

Universality requires that Efimov physics be described by the 
same parameters for large positive and negative $a$
on opposite sides of a point where $a \to \pm \infty$.  
The $^{133}$Cs experiment of Ref.~\cite{Grimm06} studied Efimov physics
for large positive and negative $a$ on opposite sides 
of the zero that comes from the interplay
between a broad Feshbach resonance near 0 magnetic field
and the large off-resonant scattering length for $^{133}$Cs.
Is there any relation 
between the Efimov parameters for the two regions on opposite 
sides of a zero of the scattering length created by a Feshbach resonance?
Is there any relation between the 
Efimov parameters associated with the large off-resonant scattering length
and the Efimov parameters associated with the 
various Feshbach resonances?
These questions can be answered by accurate measurements
of the Efimov parameters.

The Innsbruck experiment demonstrated that the resonant inelastic losses 
associated with an Efimov trimer at the 3-atom threshold
could be used to determine the Efimov parameters accurately for 
large negative $a$ \cite{Grimm06}.  
The Innsbruck group has also observed resonant inelastic losses 
associated with an Efimov trimer at the atom-dimer threshold 
\cite{FB18Santos}.  The results we have presented
should make it possible to use those results
to determine the Efimov parameters accurately for 
large positive $a$.  
These parameters could then be used to make universal predictions
for other phenomena associated with Efimov physics, such as the 
locations of local minima and maxima in the 3-body recombination rate.

We thank R.~Grimm  for valuable discussions.
This research was supported in part by the Department of Energy 
under grant DE-FG02-05ER15715.

\end{document}